\documentclass[twocolumn,pra,unsortedaddress,showpacs,floatfix,
citeautoscript,nofootinbib]{revtex4-1}
\usepackage{amsmath,amssymb,mathrsfs}
\usepackage{amsbsy}
\usepackage[usenames]{color}
\usepackage{psfrag}
\usepackage{graphicx}
\usepackage{epsfig}
\usepackage{bm}
\usepackage{color}

\usepackage[normalem]{ulem}
\usepackage{latexsym}
\usepackage[colorlinks,urlcolor=blue,citecolor=blue]{hyperref}
\usepackage{dcolumn}
\usepackage[normalem]{ulem}
\usepackage{subfigure}

\begin{document}

\title [Quasi-1D SO- and Rabi-coupled bright dipolar BEC solitons] 
{Quasi-one-dimensional spin-orbit- and Rabi-coupled bright dipolar
Bose-Einstein-condensate solitons}
\author{Emerson Chiquillo}
\address{Escuela de F\'isica, Universidad Pedag\'ogica
y Tecnol\'ogica de Colombia (UPTC),\\
Avenida Central del Norte, 150003 Tunja, Colombia.}

\begin{abstract}
We study the formation of stable bright solitons in quasi-one-dimensional
(quasi-1D) spin-orbit- (SO-) and Rabi-coupled two pseudospinor dipolar 
Bose-Einstein-condensates (BECs) of $^{164}$Dy atoms in the presence of 
repulsive contact interactions.
As a result of the combined attraction-repulsion effect of both
interactions and the addition of SO and Rabi couplings, two kinds of 
ground states in the form of self-trapped bright solitons can be formed, a
plane-wave soliton (PWS) and a stripe soliton (SS). 
These quasi-1D solitons cannot exist in a condensate with purely 
repulsive contact interactions and SO and Rabi couplings (no dipole).
Neglecting the repulsive contact interactions, our findings also show the 
possibility of creating PWSs and SSs.
When the strengths of the two interactions are close to each other, the SS
develops an oscillatory instability indicating a possibility of a breather
solution, eventually leading to its destruction.
We also obtain a phase diagram showing regions where the solution is a
PWS or a SS.
\end{abstract}


\maketitle
\section{Introduction}
In the last decade, the long-range and anisotropic dipolar interactions 
in ultracold atomic gases have provided an important environment for 
studying quantum many-particle systems \cite{Report-dip}.
Bose-Einstein condensates (BECs) of several atoms with large magnetic 
dipole moments such as chromium \cite{Chromium}, erbium \cite{Erbium}, 
and dysprosium \cite{Dysprosium} have been observed in experiments.
Recently, the experimental creation of synthetic non-Abelian gauge fields
in neutral ultracold gases \cite{SOC-1}, has opened the door to a 
fascinating and fast development of phenomena for spin-orbit- (SO-) 
coupled ultracold atoms, Dirac materials, topological insulators,
Majorana fermions, among others \cite{spin-orbit}.
In particular, within these phenomena we have the matter-wave solitons.
A fascinating well-known feature of these nonlinear waves is their
propagation without changing their shape as result of the balance 
dispersion-nonlinearity.
In BECs, the solitons have been the central focus of many works in both
condensates with dipolar interactions \cite{Report-dip}, and SO- and 
Rabi-coupled condensates, where these have given rise to a rich variety of
ground-state spin structures \cite{Solitons-1D-2D}.
Nowadays, the combined effects of spin-orbit coupling (SOC) and nonlocal
dipolar interactions have attracted great interest.
An experimental scheme to create SOC in spin-3 Cr atoms using Raman 
processes was proposed \cite{SOC-DIP-Cr}.
The combination of the Rashba SOC and the dipolar interaction predicts a
thermodynamically stable ground state with a meron spin configuration
\cite{Meron}.
In a two-dimensional (2D) SO-coupled dipolar BEC with repulsive contact 
interactions, two types of solitons have been found
\cite{Bright-solitons-2D}.
In a dipolar spin-1 BEC trapped in a double-well potential the ground 
state, the magnetic properties and the collisional and magnetic 
field quench dynamics of coupled spin-vortex pairs are investigated 
\cite{Spin-vortex}.
A spin-1 BEC with Rashba SO and dipolar interactions confined in a 
cigar-shaped trap exhibits a rich variety of ground state spin structures,
including twisted spin vortices \cite{Twisted-spin-vortices}.
Stable 2D anisotropic solitary vortices are constructed in a model of a
spinor dipolar BEC without the contact nonlinearity \cite{2D-solitons}.
In dipolar spinor BECs with SOC, gap solitons can be obtained in free 
nearly 2D space \cite{Gap-solitons}.
Cores of singular vortices with non-Abelian charges of a spinor spin-3 
$^{52}$Cr condensate are analyzed in Ref. \cite{Vortices-SOC-DIP}.
However, whether SO and Rabi coupled bright solitons can exist in 
quasi-1D dipolar BECs has not yet been investigated.

In this paper we consider a pseudospin-1/2 system in a mean-field 
treatment given by the Gross-Pitaevskii equation (GPE).
We examine the existence and properties of quasi-1D bright solitons in 
SO- and Rabi-coupled dipolar BECs of atoms of $^{164}$Dy with repulsive 
contact interactions.
While in SO- and Rabi-coupled quasi-1D condensates with repulsive contact 
interactions stable and free solitons cannot be formed, these are found 
under the combined action of the repulsion given by the contact atomic
interaction and the attraction provided by the nonlocal dipolar 
interaction.
Our findings show that the two couplings play a crucial role in the rich
ground state of the condensate. In particular, the emergence of two
different stable ground states is predicted, a PWS and a SS. 
We also  show the possibility of creating this new kind of solitons in the 
absence of the repulsive contact interactions.
Further, provided that the strengths of the two interactions are close to 
each other, the SS develops an oscillatory instability indicating a 
possibility of a breather solution, eventually leading to its destruction.
We also obtain a phase diagram showing regions where the solution is a 
plane-wave soliton (PWS) and a stripe soliton (SS).
\vskip -0.5cm
\section{Mean-field model} 
\label{II}
At ultra-low temperatures in the weakly interacting regime, properties of
a binary dipolar BEC with the dipoles oriented along the $z$ direction 
and the SO and Rabi couplings can be described by means of the scaled 
mean-field nonlocal three-dimensional (3D) GPE
\begin{widetext}
\begin{eqnarray} 
i{\partial \psi_{j}({\bf r},t)\over \partial t}  &=&
\bigg \{ -{\frac{1}{2}} \nabla^{2} + V({\bf r}) 
+(-1)^{j-1} \Big(i\gamma \frac{\partial}{\partial z} + \delta \Big)
+ 2\pi \big[g_{jj}\left|\psi_{j}({\bf r},t)\right|^2 
+ g_{12} \left|\psi_{3-j}({\bf r},t)\right|^2\big]
\nonumber \\
&+& g_{dd}\int d{\bf r'} U_{dd}(|{\bf r}-{\bf r'}|)
\big[|\psi_{1}({\bf r'},t)|^2 + |\psi_{2}({\bf r'},t)|^2 \big]
\bigg \} \psi_{j} ({\bf r},t) +\Gamma \psi_{3-j}({\bf r},t).
\label{3D-eq-dimensionless}
\end{eqnarray} 
\end{widetext}
The dimensionless form is obtained taking the harmonic oscillator (HO) 
length of the transverse trap $l_{\perp}=\sqrt{\hbar/m\omega_\perp}$, 
with the trapping frequency $\omega_\perp$.
The time $t$, the spatial variable ${\bf r}$, the energy, and the
wave functions are given in units of
$\omega^{-1}_{\perp}$, $l_{\perp}$, $\hbar \omega_{\perp}$ and
$l_{\perp}^{3/2}$, respectively.
The time-dependent spinor wave functions $\psi_{j}$ $(j=1,2)$ 
describe the two pseudospin components
$|\uparrow \rangle$ and $|\downarrow \rangle$, respectively.
Here $\int_{-\infty}^{+\infty}d{\bf r}|\psi_{j}({\bf r},t)|^2=N_{j}$,
with $N_{j}$ the number of atoms in the $j$th component, and the 
conserved total number of atoms $N=N_{1}+N_{2}$.
The external confinement potential is given as 
$V(\mathbf{r})= \rho^2/2 + V(z) $, with $\rho^2 \equiv x^2 + y^2 $.
The HO potential keeps the confinement of the system in the transverse
$(x,y)$ plane. $V(z)$ is a generic potential in the $z$ axial direction.
The strengths of the SO and Rabi couplings are 
$\gamma \equiv k_{L}l_{\perp}$ and
$\Gamma  \equiv  \Omega /(2\omega_{\perp})$, respectively,
where $k_L$ is the wave number of the Raman lasers that couple the two
atomic hyperfine states in the $z$ direction \cite{Gap-solit-OL}, and 
$\Omega$ is the frequency of the Raman coupling, responsible for the Rabi
mixing between the states.
$\delta$ is the detuning, and for simplicity, we restrict the treatment to
$\delta=0$.
The strengths of the intra- and inter-species interactions are
$g_{jj}\equiv 2\tilde{a}_{jj}/l_{\perp}=2a_{jj}$ and
$g_{12}\equiv 2\tilde{a}_{12}/l_{\perp}=2a_{12}$ with $\tilde{a}_{jj}$
and $\tilde{a}_{12}$ the respective s-wave scattering lengths.
The strength of the dipolar interaction is defined as
$a_{dd}\equiv \tilde{a}_{dd}/l_{\perp}
=\mu_0 |\boldsymbol{\mu}|^2 m/ (12\pi\hbar^2l_{\perp})$ 
and $g_{dd}=3a_{dd}$, with $\mu_0$ the permeability of free space and
$\boldsymbol{\mu}$ the dipole moment of an atom.
We have $U_{dd}(|{\bf r}-{\bf r'}|)=
(1-3\cos^2\theta)/|{\bf r}-{\bf r'}|^3$, where ${\bf r}-{\bf r'}$ 
determines the relative position of dipoles, and $\theta$ is the angle 
between ${\bf r}-{\bf r'}$ and the direction of polarization $z$.
For a strong trap in the $\rho$ direction an 1D mean-field model can be 
derived assuming that the dynamics of the condensate in the $\rho$
direction is confined in the ground state. We assume the 1D condition,
where the chemical potential satisfies $|\mu|\ll 1$ \cite{Applicability1}. 
This is tantamount to $l_{\perp}\ll \xi$ \cite{Applicability2},
with the healing length $\xi$. 
These conditions are discussed in detail later.
Thus the wave function may be split as
$\psi_{j}({\bf r},t) = \varphi_{j}(\rho)\phi_{j}(z,t)$
\cite{Dim-reduct}, where $\varphi_{j}(\rho) =  
\exp{\left(-{\rho^2 / 2\eta^2_{j}}\right)}/\sqrt{\pi}\eta_{j} $.
Using this ansatz in Eq. (\ref{3D-eq-dimensionless}) and, after 
integrating out the $\rho$ dependence, with $V(z)=0$, we get the 1D 
equation for the study of the formation of a free soliton in the $z$ 
direction
\begin{widetext}
\begin{eqnarray}
i {\partial \phi_{j}(z,t) \over \partial t}  &=&
\bigg \{ -{\frac{1}{2}} \frac{\partial^2}{\partial z^2}
+(-1)^{j-1} i \gamma \frac{\partial}{\partial z}
+ g_{jj} \left|\phi_{j}(z,t)\right|^2 
+ g_{12} \left|\phi_{3-j}(z,t)\right|^2 
\nonumber \\
&+& g_{dd} \int_{-\infty}^{+\infty} dz'
\big[V_{dd,j}(|z-z'|)|\phi_{j}(z',t)|^2
+ F_{dd,3-j}(|z-z'|)|\phi_{3-j}(z',t)|^2\big] \bigg\} \phi_{j}(z,t) 
+\Gamma \phi_{3-j}(z,t),
\label{1D-eq}
\end{eqnarray}
\end{widetext}
where $\int_{-\infty}^{+\infty}dz|\phi_{j}(z,t)|^2=N_{j}$, and
$N = N_1 + N_2$.
The integral dipolar term is evaluated in momentum space
\cite{convolution}.
For simplicity, in this paper we consider the full symmetric case:
$g_{jj}=g_{12}\equiv 2a$.
In order to obtain stationary solutions in presence of the linear coupling
provided by the Rabi term, the chemical potential must be the same for 
both components \cite{linear-coupling}. So, we construct stationary 
states setting
$\phi_{j}(z,t) \rightarrow \sqrt{N} \phi_{j}(z)\exp{(-i\mu t)}$.
The two resulting stationary equations to the fields $\phi_{1}(z)$ and
$\phi_{2}(z)$ are tantamount to each other, and these satisfy the 
condition $\phi_{1}^*(z)=\phi_{2}(z)$ \cite{Norm1,Emer}.
This condition does not apply to asymmetric solutions
\cite{Quantum-Tricriticality}.
Therefore establishing
$\phi_{1}(z) = \Phi(z)/\sqrt{2}$ and taking into account the condition 
on the wave functions we get only one stationary equation to investigate
the ground state of quasi-1D SO- and Rabi-coupled bright dipolar BEC 
solitons,
\begin{eqnarray} 
\mu \Phi &=&
\bigg [ -{\frac{1}{2}} \frac{d^2}{d z^2} + 
i \gamma \frac{d}{d z} + 2aN|\Phi|^2
\nonumber \\
&+& 3a_{dd}N\int_{-\infty}^{+\infty} dz' V_{dd}(|z-z'|)|\Phi(z')|^2\bigg]
\Phi + \Gamma \Phi^{*},
\label{Stationary-GPE}
\end{eqnarray}
with $\int_{-\infty}^{+\infty}dz|\Phi(z)|^2=1$.
To explore the above-mentioned condition $|\mu|\ll 1$ of applicability of
mean-field Gross-Pitaevskii theory to one dimension, we consider a 
homogeneous quasi-1D dipolar condensate where there is no presence of 
confinement potential neither couplings.
In this system
$\mu(a,a_{dd},n)= |\Phi|^2 \lim_{|k_z|\to 0} \tilde{V}(|k_z|)$
\cite{Applicability3}, and the interaction potential in momentum
space $\tilde{V}(|k_z|)=2aN + 4\pi N a_{dd}h(|k_z|\eta/\sqrt2)$ 
\cite{Applicability1,Dim-reduct,convolution}.
So $\mu (a,a_{dd},n) = 2n(a - a_{dd}/\eta^2)$, with the density
$n=N|\Phi|^2$. In this paper we use $\eta=1$.
Thus, the 1D mean-field regime is reached if $|a-a_{dd}|\ll 1/2n$, 
a condition satisfied in this paper.
The validity of the 1D mean-field results also can be considered requiring
that the condition $\xi/d \gg 1$ between the healing length $\xi$
and the mean interparticle separation $d=n^{-1}$ is satisfied 
\cite{Applicability4}. Here $\xi = l^2_{\perp}/(\sqrt{2} c)$ with $c$ the 
sound velocity of a homogeneous quasi-1D BEC, which can be 
read as $c=\sqrt{n\partial\mu/\partial n}=\sqrt{2n(a-a_{dd})}$.
So the condition $\xi/d \gg 1$ is equivalent to requiring 
$nl^2_{\perp}/|a-a_{dd}|\gg 1$, and also is satisfied in this paper. 
Even if strictly speaking the 1D condensation is absent, we have a 
quasicondensation, and the above discussion shows that for distances much
larger than the healing length and hence than the average interparticle 
distance, the mean-field approach is valid in dealing with such a 1D 
system \cite{Applicability4}.
In the absence of SOC $\gamma=0$, the solutions of 
Eq. (\ref{Stationary-GPE}) are real, and the resulting equation is
tantamount to the usual version of the dipolar GPE with a shifted 
chemical potential, $\mu \rightarrow \mu-\Gamma$.
In general, solutions of Eq. (\ref{Stationary-GPE}) are complex if
$\gamma \neq 0$.
For $\Gamma=0$ the SO term can be removed from Eq. (\ref{Stationary-GPE})
by substitution $\Phi(z)=\Phi_0(z)e^{i\gamma z}$, where $\Phi_0(z)$ is
the solution of a conventional stationary dipolar GPE, with a shifted
chemical potential, $\mu \rightarrow \mu + \gamma^2/2$.
Regarding the symmetry of Eq. (\ref{Stationary-GPE}), note that if
$\Gamma<0$, the restriction of the wave functions is given as
$\phi_{1}^*(z)=-\phi_{2}(z)$, which is tantamount to
Eq. (\ref{Stationary-GPE}) with $\Gamma$ replaced by $-\Gamma$.
Further, this equation is symmetric with respect to the change of $\gamma$
by $-\gamma$.

\section{Numerical results}
\label{III}
We consider the total number of atoms $N$ and the pseudomagnetization 
$M\equiv N^{-1} \sum_{j=1,2} (-1)^{j-1}N_{j}$ as constraints to compute
the ground state of SO- and Rabi-coupled quasi-1D dipolar BECs
\cite{Norm1,Norm2}.
It is important to stress that in a spinor dipolar BEC the 
pseudomagnetization is conserved under the presence of a magnetic field
\cite{magnetization1}. 
However experimentally also it is observed how dipolar interactions
induce magnetization dynamics as the magnetic field is quenched below an
extremely low value \cite{magnetization2}.
In order to find the ground state of Eq. (\ref{1D-eq}), we use a 
split-step Crank-Nicolson method with imaginary time propagation 
\cite{Norm2}.
In the imaginary time propagation $(t\rightarrow-it)$, the time evolution 
operator is not unitary, and we have conservation of neither the
normalization nor the magnetization.
To fix both the normalization and the magnetization we propose the
following approach to renormalize the wave function after each operation 
of the Crank-Nicolson method. 
We consider the continuous normalized gradient flow discussed in Ref.
\cite{Norm3}.
After each iteration the wave functions in Eq. (\ref{1D-eq})
are transformed as $\phi_{j}(z,t + \Delta t) = d_{j}\phi_{j}(z,t)$
$(j=1,2)$, where $d_{j}$ are the normalization constants.
The constraint on the total number of atoms can be written in terms
of $d_{j}$'s and the wave function components $N_{j}$ as 
$\sum_{j=1,2} d_{j}^{2}N_{j}=N$.
We have two unknowns $d_{1}$ and $d_{2}$, and only one equation, which
is given by the condition on $N$.
To find the values of the normalization constants $d_{j}$, we introduce
the constraint on the total pseudomagnetization as
$N^{-1} \sum_{j=1,2} (-1)^{j-1}d_{j}^{2}N_{j} = M$.
Solving the nonlinear system of equations given by these two conditions,
we get explicitly the normalization constants as
$d_{j}= [N[1+(-1)^{j-1}M]/(2N_j)]^{1/2}$ \cite{Norm1}.
We investigate the stationary and symmetric solutions of 
Eq. (\ref{1D-eq}), where the two internal states are equally populated.
In this case the total pseudomagnetization $M$ is zero and the total
number of atoms is $N=1$.
So the normalization constants become 
$d_{1}=d_{2}\equiv d =
(2\int_{-\infty}^{+\infty} dz |\Phi(z)|^2)^{-1/2}$.
If only the total number of atoms is considered as a conserved quantity,
our findings increase a factor of $\sqrt{2}$. However the different 
behaviors are maintained.
In this paper we consider $N=10^3$ atoms of $^{164}$Dy in the study of
SO- and Rabi-coupled bright solitons in quasi-1D dipolar BECs.
The $^{164}$Dy atoms have the largest magnetic moment of all the dipolar 
atoms used in BEC experiments.
In units of Bohr radius $a_0$, we have $^{52}$Cr ($a_{dd}\approx 15.3a_0$)
\cite{Chromium}, $^{168}$Er ($a_{dd}\approx 66.6a_0$) \cite{Erbium}, and
$^{164}$Dy ($a_{dd}\approx 132.7a_0$) \cite{Dysprosium}.
We take $l_{\perp}=\sqrt{\hbar/m\omega_{\perp}}=1\mu$m.
\begin{figure}[t] 
\begin{center}
\includegraphics[width=8.7cm,clip]{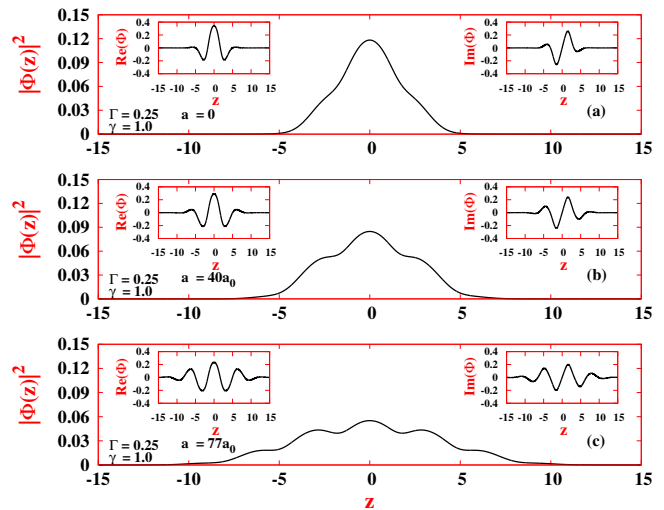}
\end{center}
\caption{(Color online)
Profiles of SO- and Rabi-coupled bright dipolar solitons for $N=10^3$ 
atoms of $^{164}$Dy ($a_{dd}\approx 132.7a_0$ in units of Bohr 
radius $a_0$). 
We set the SO and Rabi couplings $\gamma=1.0$ and $\Gamma=0.25$, 
respectively. 
From here on the couplings are used in dimensionless form.
We consider different values of the repulsive atomic interaction
(a) $a=0$, (b) $a=40a_0$ $(\approx 2.1$ nm$)$ and 
(c) $a=77a_0$ $(\approx 4.1$ nm$)$.
Inset: The real and imaginary parts of the wave function.
Lengths are measured in units of 
$l_{\perp}=\sqrt{\hbar/m\omega_{\perp}}=1\mu$m.}
\label{dip1}
\end{figure}

Figure \ref{dip1} shows the profile of SO- and Rabi-coupled bright dipolar
solitons $|\Phi(z)|^2$ for $N=10^3$ atoms of $^{164}$Dy as a function of 
axial coordinate $z$. 
We set the values of the dimensionless SO and Rabi couplings $\gamma=1.0$
and $\Gamma=0.25$, respectively, and three different values of the 
repulsive atomic contact interaction $a>0$.
The increase of $a$, causes the atoms to spread out in the condensate.
The interplay between the attractive nonlocal dipolar interaction, the
atomic short-range repulsion, and the SO and Rabi couplings gives rise to
the appearance of a new kind of stripe bright dipolar solitons. These
solitons are not present in a conventional dipolar condensate.
The rearrangement of the atoms inside the condensate results in a 
less dense soliton with expansion of its profile and an increase of the 
local maxima.
In left and right insets we also display the real and imaginary parts of
the wave function, respectively. The SS keeps the parity of the soliton 
in a condensate with SO and Rabi couplings without dipolar interaction,
where the real component has even parity while the imaginary one is odd.
It is worth noting that if the repulsive dipolar interaction $(a_{dd}<0)$
and the repulsive contact contribution are considered, the system is 
unstable and the quasicondensation is not possible.

\begin{figure}[t] 
\begin{center}
\includegraphics[width=8.7cm,clip]{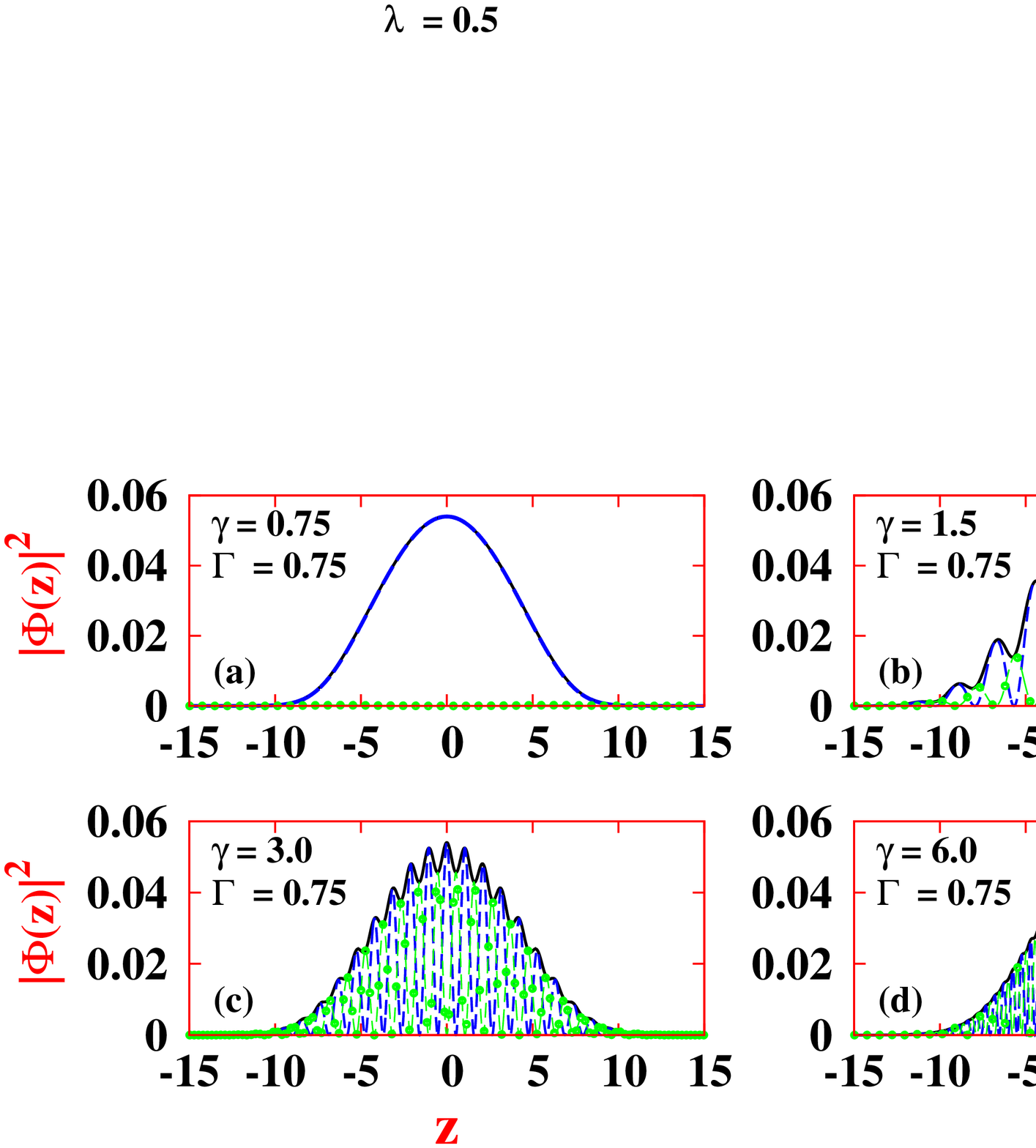}
\end{center}
\vskip -.6cm
\caption{(Color online)
SO- and Rabi-coupled bright dipolar solitons. Here $N=10^3$ atoms of
$^{164}$Dy with $a=77a_0$, $\Gamma=0.75$ and four different values of 
$\gamma$.
The solid line depicts $|\Phi(z)|^2$, and the dashed and dotted lines
represent squared real and imaginary parts of the wave function, 
respectively.
Lengths are measured in units of $l_{\perp}$.}
\label{dip2}
\end{figure}

Next we consider the effect of $\gamma$ and $\Gamma$ on bright dipolar
solitons.
In Fig. \ref{dip2} the numerical results are reported for a condensate of
$N=10^3$ atoms of $^{164}$Dy with $a=77a_0$, the Rabi coupling
$\Gamma=0.75$, and four different values of the SO coupling $\gamma$.
In Fig. \ref{dip2} (a) we have a PWS, where the imaginary component of the
wave function becomes very small.
At fixed $\Gamma$, the increase of $\gamma$ provides a linkage between the
atoms, and it leads to the wave function showing oscillatory real and 
imaginary components with an increase of the imaginary part. As a result 
the profile develops several local maxima giving rise to a multi-peak
nature of the soliton or stripe bright dipolar soliton [Figs. 
\ref{dip2}(b) and \ref{dip2}(c)].
At large values of $\gamma$ [Fig. \ref{dip2} (d)], we have many 
small-amplitude local variations of the density, which makes the averaged 
profile of the soliton close to that of a PWS.
It is worth noting that the profile and the size of the solitons remain
almost constant because of the combined attraction-repulsion effect of 
the interactions, contrary to the usual bright solitons in SO- and 
Rabi-coupled condensates \cite{Emer}.

\begin{figure}[t] 
\begin{center}
\includegraphics[width=8.7cm,clip]{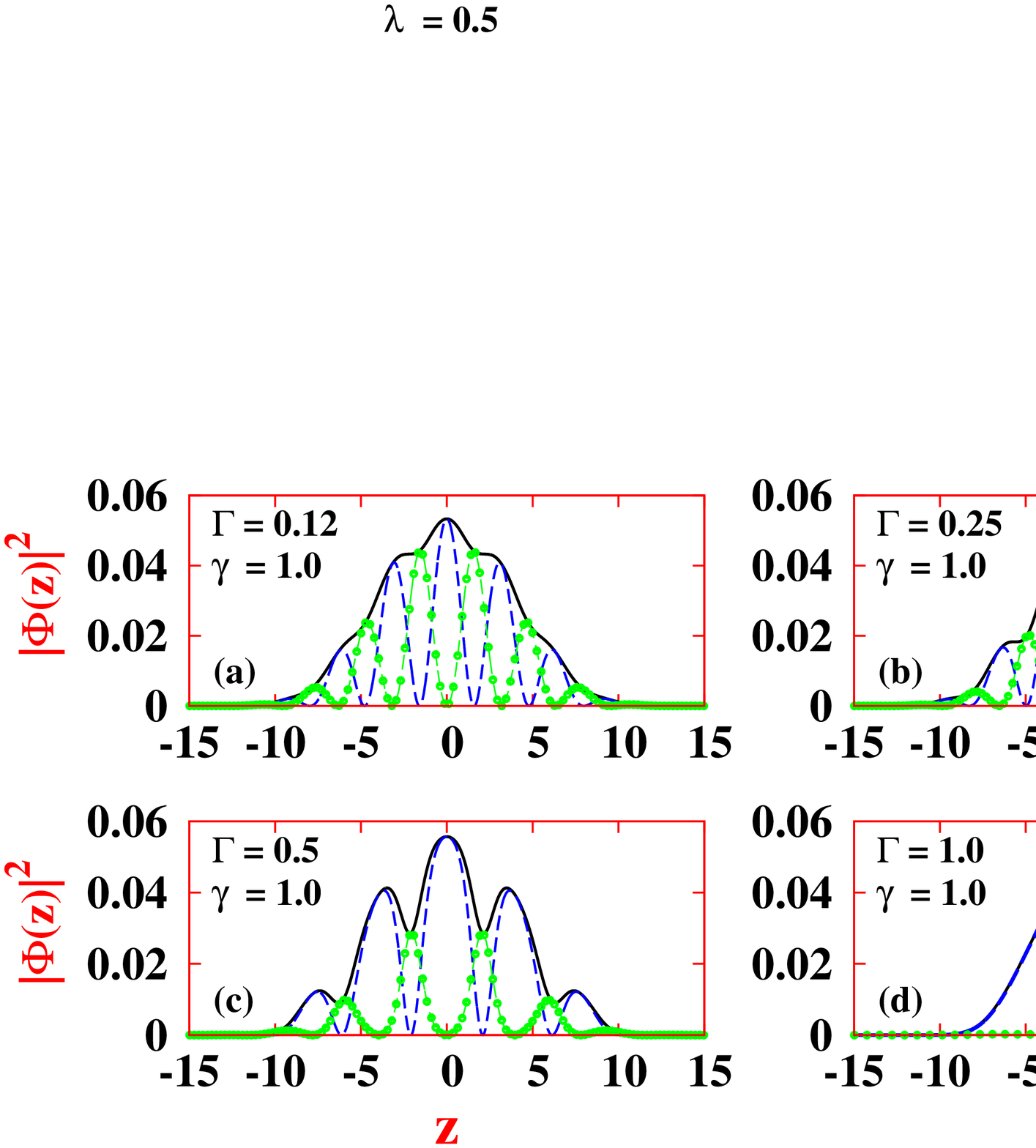}
\end{center}
\vskip -0.6cm
\caption{(Color online) The same as in Fig. \ref{dip2}, but for a fixed
SOC $\gamma=1.0$, and four values of the Rabi coupling $\Gamma$.}
\label{dip3}
\end{figure}

It is also relevant to analyze the bright dipolar solitons at fixed 
$\gamma$ and different values of $\Gamma$ as plotted in Fig. \ref{dip3},
where we have $N=10^3$ atoms of $^{164}$Dy with
$a=77a_0$, $\gamma=1.0$, and four values of $\Gamma$.
In this case our findings show that the increase of mixing between states,
accounting for $\Gamma$, leads to the decrease of the imaginary part
of the wave function with a slow increase in the number of local maxima.
When $\Gamma$ is sufficiently large, the imaginary component of the 
density is almost suppressed as is seen in the Fig. \ref{dip3} (d)
and the SS being nearly identical to a PWS.
\begin{figure}[t] 
\begin{center}
\includegraphics[width=8.7cm,clip]{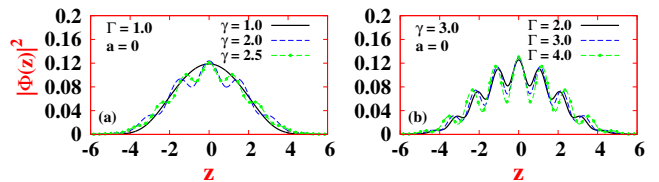}
\end{center}
\vskip -0.6cm
\caption{(Color online) SO and Rabi coupled condensates of $N=10^3$ atoms
of $^{164}$Dy without contact interaction $(a=0)$.
(a) $\Gamma=1.0$, $\gamma=1.0$, $2.0$ and $2.5$.
(b) $\gamma=3.0$, $\Gamma=2.0$, $3.0$ and $4.0$.
Lengths are measured in units of $l_{\perp}$.}
\label{dip4}
\end{figure}

In Fig. \ref{dip4} we analyze SO- and Rabi-coupled dipolar BECs in the 
absence of the atomic contact interaction $(a=0)$.
A noteworthy feature is a shrinking, high, and constant soliton profile.
In Fig. \ref{dip4} (a) we set $\Gamma=1$, and with the increase of 
$\gamma$ we found that a PWS undergoes a transition to a SS, which has a
slow increase of the density oscillations compared to Fig. \ref{dip2}.
In Fig. \ref{dip4} (b) we set $\gamma=3$ and different values of 
$\Gamma$. In this case we do not have a PWS, there are only SSs.
The increase of $\Gamma$ causes a reduction of the imaginary part of the
wave function in a similar way as is described in Fig. \ref{dip3}, but
without a change in the number of oscillations nor suppression of the 
SS.
Results of Fig. \ref{dip4} demonstrate that the interplay of the two 
couplings $(\gamma,\Gamma)$, the nonlocal dipolar interaction, and the 
contact interaction produces new features for SO- and Rabi-coupled dipolar
BECs.
\begin{figure}[t] 
\begin{center}
\includegraphics[width=8.7cm,clip]{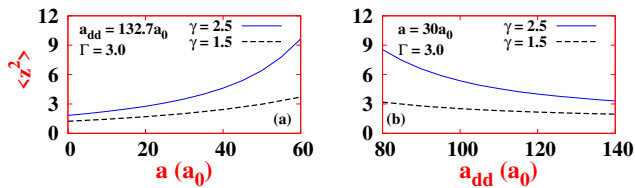}
\end{center}
\vskip -0.6cm
\caption{(Color online) RMS size of the SO- and Rabi-coupled dipolar BEC 
solitons versus (a) the contact interaction $a$, with $a_{dd}=132.7a_0$,
and (b) the dipolar length $a_{dd}$, with $a=30a_0$ $(\approx 1.6$ nm$)$.
Here, $N=10^3$ atoms, $\Gamma=3.0$  and $\gamma=1.5,2.5$.
Lengths are measured in units of $l_{\perp}$.}
\label{dip5}
\end{figure}

Our aim is investigate effects of $a$ and $a_{dd}$ on the root mean
square (rms) of the self-trapped nonlinear waves. 
This trend is plotted in Fig. \ref{dip5}. 
In order to have stable bright dipolar solitons, one should have 
$a_{dd}>a>0$, where the dipolar attraction dominates over the sizable 
contact repulsion.
We results show that there exist PWSs and SSs to $\gamma=1.5$ and
$\gamma=2.5$, respectively.
In Fig. \ref{dip5} (a) we set the dipolar strength of the $^{164}$Dy 
atoms, $a_{dd}=132.7a_0$, $\Gamma=3.0$ and two values of $\gamma$.
Here the PWSs are stable. However, for $\gamma=2.5$, while $a$ approaches
$ a_{dd}$, the amplitude of the SSs develop an oscillatory instability, 
indicating a possibility of a breather solution \cite{Gap-solitons},
and eventually leading to the destruction of these.
In Fig. \ref{dip5} (b) we set the contact interaction $a=30a_0$, 
and the same values of the two couplings as in the Fig. \ref{dip5} (a).
Both PWSs and SSs are stable. 
With the attractive contribution of the dipolar term, the density profile 
shrinks, and it is reflected as a reduction in the rms size of the 
soliton.
\begin{figure}[t] 
\begin{center}
\includegraphics[width=6cm,height=4cm,clip] {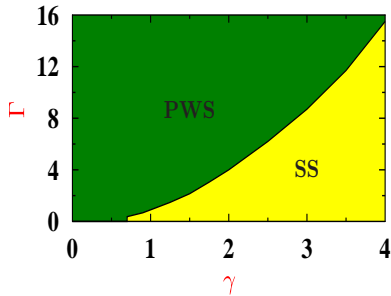}
\end{center}
\vskip -0.6cm
\caption{(Color online) $\Gamma-\gamma$ phase diagram of SO- and 
Rabi-coupled bright dipolar solitons of $N=10^3$ atoms of $^{164}$Dy with
$a=50a_0$ $(\approx 2.7$ nm$)$.}
\label{dip6}
\end{figure}

In Fig. \ref{dip6} we obtain a $\Gamma-\gamma$ phase diagram of $N=10^3$ 
atoms of $^{164}$Dy with $a=50a_0$. We show the regions where the profile 
is a PWS and a SS.
For $\gamma < 0.7$ and any finite value of $\Gamma$ we have only PWSs.
When $\gamma \geq 0.7$ and the corresponding $\Gamma$, the phase is a SS.
If $\Gamma$ tends to zero in Eq. (\ref{Stationary-GPE}), we do not have
the mixing between states, and the solutions of this equation are PWSs. 
We find a small region for $\Gamma \leq 0.03$ and 
$0.7 \leq \gamma \leq 4.0$ where it is possible to have PWSs, though it is
not included in the phase diagram.

\section{Summary and outlook}
We have investigated the possibility of creating stationary and stable
bright solitons in SO- and Rabi-coupled dipolar BECs in the weakly 
interacting regime using a 1D mean-field nonlocal GPE in the $z$ axis.
The analysis was focused on effects caused by the SO and Rabi couplings in
the ground state of a dipolar condensate of $^{164}$Dy atoms.
The numerical results demonstrate interesting changes of the soliton
profile due to the interplay of SO and Rabi couplings.
Both, the absence and the presence of the repulsive contact interaction
effects added to the dipolar interaction were studied. 
Two kinds of stable SO- and Rabi-coupled bright dipolar solitons, PWSs and
SSs, were found.
The solitons persist and remain stable in the presence of the local 
self-repulsion, balanced by the effectively attractive dipolar 
contribution.
When the strengths of the two interactions are close to each other, the SS
develops an oscillatory instability, indicating a possibility of a
breather solution and eventually leading to its destruction.
These results provide exciting new insight into the nonlinear physics of 
dipolar BECs in artificially induced gauge fields.
An interesting possibility is to extend the present analysis from solitons
to quantum droplets \cite{Applicability1,LHY}, where is revealed the
crucial role played by the first correction beyond a mean-field
approximation.
Other important questions, such as the stability analysis of PWSs and SSs
under the action of a trap potential, remain to be investigated. 
In the absence of the dipolar interactions this question was 
studied in Refs. \cite{Emer,Trapped}.
Although our discussion is focused in the context of BECs, similar
physics can be investigated in other physical systems which are governed
by coupled GPEs, such as fiber optics \cite{fiber-optics}, 
exciton-polaritons \cite{exciton-polaritons}, quantum phase transitions
\cite{quantum phase transitions}, SO-coupled Mott insulators, and 
superfluids of atoms in optical lattices \cite{Mott insulators}, 
among others.

\end{document}